\documentclass[useAMS,usenatbib]{mnras}
\usepackage{graphicx}
\usepackage{epstopdf}
\usepackage{amsmath,esint,bm}
\usepackage{amssymb}

\usepackage{etoolbox}
\makeatletter
\patchcmd\@combinedblfloats{\box\@outputbox}{\unvbox\@outputbox}{}{%
   \errmessage{\noexpand\@combinedblfloats could not be patched}%
}%
 \makeatother

\title[Universal expansion with spatially varying $G$]{Universal expansion with spatially varying $G$}
\author[Christodoulou \& Kazanas]{Dimitris M. Christodoulou$^{1,2}$\thanks{This work is dedicated to the memory of Ren\'e Thom whose ``Catastrophe Theory'' gave us astonishing solutions in various astrophysical settings starting back in 1995 \citep{chr95a,chr95b}, and it continues to be relevant to this date and in this investigation.}  and Demosthenes Kazanas$^{3}$
\\
$^{1}$Lowell Center for Space Science and Technology, University of Massachusetts Lowell, Lowell, MA, 01854, USA.\\
$^{2}$Dept. of Mathematical Sciences, Univ. of Massachusetts Lowell,
Lowell, MA, 01854, USA. E-mail: dimitris\_christodoulou@uml.edu\\
$^{3}$NASA/GSFC, Laboratory for High-Energy Astrophysics, Code 663, Greenbelt, MD 20771, USA. E-mail: demos.kazanas@nasa.gov\\
}

\begin{document}

\def\gsim{\mathrel{\raise.5ex\hbox{$>$}\mkern-14mu
                \lower0.6ex\hbox{$\sim$}}}

\def\lsim{\mathrel{\raise.3ex\hbox{$<$}\mkern-14mu
               \lower0.6ex\hbox{$\sim$}}}

%\date{Accepted ??? . Received 2017 March ??; in original form 2017 March ??}
\pagerange{\pageref{firstpage}--\pageref{lastpage}} \pubyear{2019}

\maketitle

\label{firstpage}

\begin{abstract}
We calculate the expansion of the universe under the assumptions that $G$ varies in space and the radial size $r$ of the universe is very large (we call this the MOND regime of varying-$G$ gravity). The inferred asymptotic behavior turns out to be different than that found by McCrea \& Milne in 1934 and our equations bear no resemblance to those of the relativistic case. In this cosmology, the scale factor $R(t)$ increases linearly with time $t$, the radial velocity is driven by inertia, and gravity is incapable of hindering the expansion. Yet, Hubble's law is borne out without any additional assumptions. When we include a repulsive acceleration $a_{\rm de}$ due to dark energy, the resulting universal expansion is then driven totally by this new term and the solutions for $a_{\rm de}\to 0$ do not reduce to those of the $a_{\rm de}\equiv 0$ case. This is a realization of a new Thom catastrophe: the inclusion of the new term destroys the conservation of energy and the results are not reducible to the previous case in which energy is conserved.
\end{abstract}




\section{Introduction}

In the FLRW relativistic metric of the general relativity \citep{wei72,kaz80,ish19,fer19} as well as in the Newtonian cosmology \citep{mcc34,mil35,gur85}, the radial expansion of the scale of the spherical universe $R(t)$ is described by the following differential equation in the absence of dark energy:
\begin{equation}
\left(\frac{\dot{R}}{R}\right)^2 = \frac{8\pi G_0}{3}\rho -  \frac{k c^2}{R^2}\, ,
\label{newt}
\end{equation}
where $G_0$ is the Newtonian gravitational constant, $c$ is the speed of light, $\rho(t)$ is the spatially uniform density of the medium, the dot denotes derivative with respect to cosmic time $t$, and $k$ is the curvature of space. It is rather odd that Newtonian dynamics and general relativity both lead to the same equations for the expanding universe. One reason (perhaps the only reason) for such confluent descriptions is the assumption of the same constant $G_0$ in both theories, in conjunction with the cosmological principle \citep{ber76}. We have surmised that after we solved the problem of the Newtonian universal expansion with varying $G(r)$, where $r$ is the radial coordinate. Furthermore, \cite{per19} used a Yukawa parameterization of varying-$G$ gravity and they obtained yet a different set of dynamical equations for the expansion of the universe (their equations 2.21 and 2.22). Thus, it seems that the assumption of a constant $G_0$ is too restrictive and binding in current theories of gravity; and that any variation of $G$ produces new physical models. In varying-$G$ models, as was also pointed out by the referee, the spatial variation of $G(r)$ is permissible because the center is the location of any point-mass in the universe. The superposition of all point-masses over the entire universe will lead to a homogeneous $G$.

In our spatially varying-$G$ gravity \citep{chr18,chr19}, $G$ is given by the equation
\begin{equation}
G(s(r)) =\frac{G_0}{2}\left(1 + \sqrt{1 + \frac{4}{s}} \right) \, ,
\label{eq2}
\end{equation}
where $s(r)\equiv\sigma/\sigma_0$ is the dimensionless surface density of a spherical mass distribution $M(r)$, $\sigma = M/r^2$ (where $\sigma\to 0$ as $r\to\infty$), and $\sigma_0 = a_0/G_0$. Here $a_0$ is the familiar MOND acceleration of about $1.2\, {\rm \AA}$\,s$^{-2}$ \citep{mil83,mil15,san02}. Using the above prescription for $G(s(r))$ results in cosmological equations that are not manageable analytically. We can however solve analytically for the two asymptotic cases of the Newton-Weyl regime ($s\to\infty$) and the MOND regime ($s\to 0$).

When $s\to\infty$, then equation~(\ref{eq2}) reduces to $G=G_0$ and the Newtonian treatment of \cite{mcc34} and \cite{mil35} is fully recovered. On the other end, when $s\to 0$, then equation~(\ref{eq2}) reduces to
\begin{equation}
G(s(r)) =\frac{G_0}{\sqrt{s}} = \left(\frac{3 G_0 a_0}{4\pi\rho(t) r}\right)^{1/2}\, ,
\label{eq3}
\end{equation}
to leading order in $1/s$, where we also used the definition $M(r, t)\equiv 4\pi r^3\rho(t)/3$ and the assumption that the spatially uniform density $\rho$ is only a function of time $t$ \citep[as in the study of][]{mcc34}. In this case, the cosmological principle \citep{wei72} is still valid, but the universal expansion of the scale factor $R(t)$ at late times changes its evolution and its properties dramatically as compared to the standard \cite{mcc34} Newtonian cosmology, as we describe in \S~\ref{sec2} below. In \S~\ref{sec3}, we include dark energy in the calculations and the expansion of the universe changes character and properties once again. Following these analyses, we summarize and discuss our results in \S~\ref{sum}.

\section{Universal Expansion with Varying $G$ in the MOND Asymptotic Regime}\label{sec2}

\subsection{Preliminaries}

We work in the deep MOND regime of varying-$G$ gravity, where $s\to 0$ and also $r \gg r_M$, where $r_M$ is MOND's scale length defined by the equation\footnote{Adopting fiducial values for the mass $M_{\rm U}=4.5\times 10^{51}$~kg and the radius $r_{\rm U}=4.4\times 10^{26}$~m of the observable universe and also $a_0=1.2\times 10^{-10}$~m~s$^{-2}$, we estimate that $r_M = 5.0\times 10^{25}$~m or $r_{\rm U}/r_M \simeq 9$; thus the present universe (observable and beyond) appears to be already in the MOND asymptotic regime. This is also corroborated independently by the characteristic time $T_0$ for the universe to enter the MOND regime, viz., $T_0=c/(2\pi a_0)\simeq 12.6$~Gyr \citep{mil15} which is somewhat shorter than the age of the universe ($\simeq 14$~Gyr).\label{ftu}}
\begin{equation}
r_M\equiv \sqrt{\frac{G_0 M(r)}{a_0}}\, ,
\label{rm}
\end{equation}
where mass $M(r)$ is constant within radius $r$ in a Lagrangian framework in which we move along with a test particle that is located on the surface of a uniform sphere of density $\rho(t)$. In the following, we will pursue an Eulerian description of the equations of motion and continuity, in which case, $M(r, t) =  4\pi r^3\rho(t)/3$ \citep[as in the study of][]{mcc34}.
Then equation~(\ref{eq3}) is applicable and the radial gravitational acceleration $a$ on the surface of an expanding sphere of radius $r$ is given by the equation
\begin{equation}
a\equiv \frac{G(s(r)) M(r)}{r^2} = \sqrt{{\cal A}_0\,\rho(t) r}\, ,
\label{grav}
\end{equation}
where MOND's fundamental constant ${\cal A}_0$ \citep{mil15,chr18} is defined here after absorbing a factor of $4\pi/3$ in it, viz.
\begin{equation}
{\cal A}_0\equiv \frac{4\pi}{3}G_0 a_0\, .
\label{A0}
\end{equation}
We note that $\rho r = 3\sigma/(4\pi)$ for a spherical mass distribution in Eulerian coordinates, and then equation~(\ref{grav}) can be cast in the form 
$$a = \sqrt{G_0 a_0\sigma(r)}\, ,$$
which reveals that the radial acceleration of the test particle on the surface of the sphere is uniquely determined by the surface density $\sigma(r)=M/r^2$. This is a fundamental property of varying-$G$ gravity \citep[see][]{chr19}.

\subsection{Equations of Motion and Continuity and Hubble's Law}\label{mcc34}

Following \cite{mcc34} and \cite{mil35}, the Eulerian equation of motion of a test particle at radius $r$ with speed $v$ is
\begin{equation}
\frac{\partial v}{\partial t} + v\frac{\partial v}{\partial r} = - \sqrt{{\cal A}_0\,\rho(t)\, r}\, ,
\label{motion}
\end{equation}
and the Eulerian equation of continuity is
\begin{equation}
\frac{1}{\rho}\frac{d\rho}{dt} + \frac{1}{r^2}\frac{\partial}{\partial r}\left(r^2 v\right) = 0\, .
\label{cont}
\end{equation}
Here we wrote the derivative of the density as $d\rho/dt$ because the uniform density $\rho$ of the spherical mass distribution is assumed to be a function of time only, whereas the radial speed is $v(r, t)$. Then, we set $(1/\rho)(d\rho/dt) = -3 H(t)$ and $(1/r^2)\partial(r^2 v)/\partial r = +3 H(t)$, where the function $H(t)$ is to be determined. The former equation implies that the cosmological principle remains valid, precisely as in the \cite{mcc34} study.
Integrating the latter equation, we find that
\begin{equation}
v = r\left(H(t) + \frac{J(t)}{r^3}\right)\, ,
\label{v}
\end{equation}
where $J(t)$ is the constant of integration in $r$, generally a function of $t$. Substituting equation~(\ref{v}) into equation~(\ref{motion}), we find that
\begin{equation}
\sqrt{r}\left[ \dot{H} +\frac{\dot{J}}{r^3} + \left( H +\frac{J}{r^3} \right) \left( H - \frac{2J}{r^3}\right) \right] = - \sqrt{{\cal A}_0\,\rho(t)}
\, ,
\label{big}
\end{equation}
where  the dots denote derivatives of $H(t)$ and $J(t)$ with respect to cosmic time $t$. This is the point where our analysis deviates from the calculation of \cite{mcc34}. The right-hand side of equation~(\ref{big}) is a function of time only, and the same condition applied to the left-hand side is supposed to determine the integration constant $J(t)$, which turns out to be zero in the \cite{mcc34} treatment, but not in our analysis.

Unable to determine $J(t)$, we proceed as follows: We reduce equation~(\ref{big}) to the deep MOND limit $r\gg r_M$. Then all terms with powers of $1/r^3$ can be discarded and we find that
$$\dot{H} + H^2 = -\sqrt{\frac{{\cal A}_0\,\rho(t)}{r}}\, ,$$
or, asymptotically as $r\to\infty$, that
\begin{equation}
\dot{H} + H^2 = 0\, .
\label{f}
\end{equation}
In equation~(\ref{v}), we also have to drop the $J(t)/r^3$ term for consistency, and then the expansion speed assumes the asymptotic form
\begin{equation}
v(r, t) = r\,H(t)\, .
\label{v2}
\end{equation}
Equations~(\ref{f}) and~(\ref{v2}) are fundamental for the cosmology with varying $G$ in the MOND asymptotic limit of $r\gg r_M$. Equation~(\ref{v2}), in particular, is a realization of Hubble's law which becomes valid in the MOND regime. 

On the other hand, Hubble's law is not valid in the regime of intermediate accelerations between the Newton-Weyl and MOND limits (equation~(\ref{v}) in the case of $r\sim r_M$). This is an unexpected result; it shows that Hubble's law in the present universe happens to be valid only because the universe has entered the MOND regime (see footnote~\ref{ftu}). Therefore, Hubble's law starts out to be true in the early Newton-Weyl universe ($G=G_0$), then it becomes invalid at intermediate accelerations, and finally it is reinstated in the MOND regime described by equations~(\ref{eq3}) and~(\ref{v2}).

\subsection{Varying-$G$ Cosmology in the MOND Regime}

It is not surprising that the analysis of universal expansion in the deep MOND regime $r\gg r_M$ is considerably simpler than the \cite{mcc34} treatment since we can solve equations~(\ref{f}) and~(\ref{v2}) rather easily. Before we do so, we should draw attention to the fact that the gravitational acceleration term ($\sqrt{{\cal A}_0\rho r}$) has dropped out of equation~(\ref{f}). An immediate consequence is that the expansion at late cosmic times (when $r\to\infty$ effectively) is not retarded significantly by gravitational attraction. This is not surprising: we are working in the asymptotic limit of $r\gg r_M$, thus gravity has weakened considerably and it is incapable of providing any substantial resistance to the continuing radial expansion. In the present context, there is only one other factor that can drive the expansion unimpeded by gravity, namely the inertia of the expanding spherical mass. We show this to be true in equation~(\ref{v3}) below.

The general solution of equation~(\ref{f}) is
\begin{equation}
H(t) = \frac{1}{c_1 + t}\, ,
\label{fsol}
\end{equation}
where $c_1$ is the integration constant. Substituting this solution into the equation $(1/\rho)(d\rho/dt) = -3 H(t)$ and integrating, we find that
\begin{equation}
\rho(t) = \frac{c_2}{(c_1 + t)^3}\, ,
\label{rhosol}
\end{equation}
where $c_2$ is another integration constant, and that
\begin{equation}
H(t) = \left( \frac{\rho(t)}{c_2} \right)^{1/3}\, .
\label{frho}
\end{equation}
Combining equations~(\ref{v2}) and~(\ref{frho}), we find for the expansion speed that
\begin{equation}
v = \left( \frac{3}{4\pi c_2} M(r) \right)^{1/3}\propto \,r\, .
\label{v3}
\end{equation}
This relation shows that, at late times, expansion is driven by the inertia of $M(r)$ and not by gravity. It is also notable that, in this context, the radial speed scales with mass as 
$$v^3\propto M\, .$$ 
This proportionality (which is effectively Hubble's law) should be contrasted to the Tully-Fisher and Faber-Jackson relations $v^4\propto M$ \citep{tul77,fab76} for the asymptotic rotation and dispersion velocities of spiral and elliptical galaxies, respectively \citep[see also recent works by][]{mcg00,san02,mcg12,den15}.

\subsection{Late Evolution of the Scale Factor}

Equation~(\ref{v2}) takes the form
\begin{equation}
\frac{dr}{dt} = r\,H(t)\, .
\label{v4}
\end{equation}
Integrating this equation yields the relation $r(t)\propto R(t)$, where $R(t) = ({\rm const.})\exp\left({\int{H(t) dt}}\right)$ is defined as the scale factor of the expansion of the universe. Then equation~(\ref{v4}) can be rewritten in the form
\begin{equation}
H(t) = \frac{1}{R}\frac{dR}{dt}\, ,
\label{v5}
\end{equation}
that describes the evolution of the scale factor $R(t)$.
Substituting this equation into equation~(\ref{f}), we find for the cosmological scale factor that
\begin{equation}
\ddot{R} = 0 \ \Longrightarrow \ R(t) = c_3\, t + c_4\, ,
\label{bigR}
\end{equation}
where $c_3$ and $c_4$ are integration constants. In this cosmology, the scale factor $R(t)$ will increase linearly with time $t$ at late times and at very large radii $r$. Based on the above results, it is rather obvious that inertia (eq.~(\ref{v3})) is incapable of producing a faster (e.g., exponential) expansion of the universe at late times. Nevertheless, the expansion will continue to proceed at linear rates as $t, r\to\infty$ (in the deep MOND regime of varying-$G$ gravity in the absence of dark energy).

\section{Inclusion of a Repulsive Dark-Energy Term}\label{sec3}

\cite{mcc34} claimed that the inclusion of a repulsive dark-energy term in Newtonian cosmology was ad-hoc, and this presumption is wide-spread to date. But this is no longer the case: \cite{gur85} showed that in the case of spherical symmetry the most general force law at the surface of a self-gravitating sphere is a linear combination of the Newtonian force and a Hooke-type repulsive linear force. This result justifies the addition of a repulsive dark-energy term in the Newtonian equation of motion of a spherical fluid. We also note that these two force components are known individually to be the only ones that support closed orbits, as Isaac \cite{new87} has already proven in {\it Principia}, but their linear combination has not been investigated in detail yet \cite[see however the study of][]{bar96}.

Equation~(\ref{motion}) with a Gurzadyan repulsive dark-energy term \citep{gur85,gur19,bar96,gur19b} then reads
\begin{equation}
\frac{\partial v}{\partial t} + v\frac{\partial v}{\partial r} = - \sqrt{{\cal A}_0\,\rho\, r} + \ell^2 r\, ,
\label{motion2}
\end{equation}
where we wrote Einstein's usual cosmological constant $\Lambda /3$ as $\ell^2$ with $\ell>0$ to reinforce its positive value. The continuity equation~(\ref{cont}) remains the same. Substituting equation~(\ref{v}) into equation~(\ref{motion2}), we find that
\begin{equation}
\dot{H} +\frac{\dot{J}}{r^3} + \left( H +\frac{J}{r^3} \right) \left( H - \frac{2J}{r^3}\right) = \ell^2 - \sqrt{\frac{{\cal A}_0\,\rho}{r}} \, .
\label{big2}
\end{equation}
In the deep MOND limit $r\gg r_M$, the $1/r$ terms can all be discarded and this equation reduces to
\begin{equation}
\dot{H} + H^2 = \ell^2\, ,
\label{f2}
\end{equation}
whereas equation~(\ref{v2}) for $v(r, t)$ is still valid. The general solution of equation~(\ref{f2}) is
\begin{equation}
H(t) = \ell\,\tanh\left[ \ell(c_1 + t) \right]\, ,
\label{fsol2}
\end{equation}
where $c_1$ is the integration constant. Substituting this solution into the equation $(1/\rho)(d\rho/dt) = -3 H(t)$ and integrating, we find that
\begin{equation}
\rho(t) = c_2\,{\rm sech}^3\left[ \ell(c_1 +t) \right]
\, ,
\label{rhosol2}
\end{equation}
where $c_2$ is another integration constant, and that
\begin{equation}
H(t) = \ell\,\left[ 1 - \left(\frac{\rho}{c_2}\right)^{2/3} \right]^{1/2} \, .
\label{frho2}
\end{equation}
We see that the Hubble constant becomes a true constant, equal to $\ell$, only for $\rho(t)\to 0$. Thus, in this case, the Hubble constant $H\to\ell$ is determined exclusively by dark energy.

Combining equations~(\ref{v2}) and~(\ref{frho2}), we find for the radial expansion speed that
\begin{equation}
v = \ell\,\,\sqrt{r^2 - \left(\frac{3M}{4\pi c_2}\right)^{2/3}}\, .
\label{v11}
\end{equation}
In this description, inertial resistance (the term $\propto M^{2/3}$) opposes the expansion (as inertia should) which is now driven entirely by dark energy (through the constant $\ell$).

Finally, the growth of the scale factor $R(t)$ is no longer linear in the presence of dark energy. Equations (\ref{v5}) and (\ref{f2}) imply for $R(t)$ that
\begin{equation}
\ddot{R} = \ell^2\,R \ \Longrightarrow \ R(t) = c_3\,\exp{(\ell t)} + c_4\,\exp{(-\ell t)}\, ,
\label{bigR2}
\end{equation}
where $c_3$ and $c_4$ are integration constants. As $t\to\infty$, and since $\ell > 0$, we find that
\begin{equation}
R(t) \simeq c_3\,\exp{(\ell t)}\,, \ \ \ (c_3>0)  \, .
\label{bigR3}
\end{equation}
In this cosmology that is driven by dark-energy repulsion, the scale factor $R(t)$ increases exponentially with time $t$ at late times. This is not an unexpected result. Apparently, dark energy wins against both gravitational resistance and inertial resistance at very late times (as $t, r\to\infty$) and it drives the universal expansion at an exponential rate.

\section{Summary and Discussion}\label{sum}

Using spatially varying-$G$ gravity \citep{chr18,chr19}, we have analyzed two expanding universes in the asymptotic MOND limit (at late cosmic times and at large radial sizes), one devoid of dark energy (\S~\ref{sec2}) and another with the inclusion of repulsive dark energy (\S~\ref{sec3}). In both cases, gravity is too weak to influence/retard the expansion which is driven solely by inertia or dark energy, respectively. Inertia is also weak going against dark energy in the latter case (equation~(\ref{v11})). In both cases, however, Hubble's law (equation~(\ref{v2})) remains valid without the need of resorting to additional assumptions. However, Hubble's constant becomes a true constant, independent of time, only in the dark-energy dominated universe (equation~(\ref{frho2}) with $\rho\to 0$) at late cosmic times.

A striking difference between the two cases is that the inertia can produce only a linearly expanding scale factor at late times (equation~(\ref{bigR})), whereas dark energy can produce an exponential growth of the scale factor (equation~(\ref{bigR3})). In particular:
\begin{itemize}
\item[(a)] The linear growth of the scale factor at late times in the absence of dark energy (\S~\ref{sec2}) also implies that the expansion speed scales as $v\propto M^{1/3}\propto r$. This relation shows that inertia alone drives the expansion of this universe in the absence of dark energy or any other extraneous factors.  \\
\item[(b)] The exponential growth of the scale factor at late times with the inclusion of dark energy  (\S~\ref{sec3}) implies that the asymptotic radial speed $v \simeq \ell r$ as $r\to\infty$, where $\ell = \sqrt{\Lambda/3}> 0$ is the coefficient introducing the dark-energy repulsion \citep[as in the Newtonian description of][]{gur85}. In this case, it is the dark-energy repulsion that drives the expansion unhindered by gravitational resistance and also by inertial resistance.
\end{itemize}
In both of the above cases, the asymptotic radial velocities can only be strictly positive. This leads to hyperbolic expansions in universes that can be interpreted as having negative curvatures \citep[see also][for additional cases that do not materialize in varying-$G$ gravity]{mcc34}. Furthermore, to the proponents of exponential universal expansion, the above two different outcomes would seem to support the presence of dark energy in this Newtonian universe because only then can the varying-$G$ model achieve exponential growth in time (equation~(\ref{bigR3})). On the other hand, to the extent that observations will continue to support the strange present coincidence $R_h= c\, t$ \citep{mel18,mel19}, where $R_h$ is the apparent horizon of the universe, the varying-$G$ solution without dark energy is strongly favored since the local flatness theorem \citep{wei72} is automatically satisfied and the theory predicts quite naturally that $H(t)=1/t$ (for $c_1=0$) and $R(t)=c_3\, t$ (for $c_4=0$); and it also validates the above equation for $R_h$ at present and at all future times.

The results described in \S~\ref{sec2} and in \S~\ref{sec3} also reveal the presence of a {\it Thom catastrophe} \citep{tho75,gil81} among universes with and without dark energy. If we set $\ell = 0$ in equation~(\ref{f2}), we recover equation~(\ref{f}), so the differential equations appear to behave according to our expectations. But the solutions of \S~\ref{sec3} do not reduce to those of \S~\ref{sec2} for $\ell \to 0$. In fact,  equations~(\ref{fsol2})-(\ref{bigR3}) in \S~\ref{sec3} all reduce to zero or a constant for $\ell = 0$. Thus, the results of \S~\ref{sec2} cannot be recovered from the dark-energy case as we let $\ell\to 0$. This is the signature of one of Thom's catastrophes \citep{tho75}. The phenomenon occurs when different conservation laws are applicable in the two cases \citep{chr95a,chr95b}. Once an additional conservation law is imprinted or is destroyed between the two cases, the solutions can no longer be reduced from one case to the other in the naive way expected by simple continuity arguments. Continuity is also destroyed by the application of differing conservation laws between the two states under consideration, and a nonlinear Thom catastrophe then sets in. 

In the present case, the inclusion of dark energy in \S~\ref{sec3} alters the conservation of kinetic energy as this materializes in the treatment of \S~\ref{sec2}. With gravity incapable of competing in either case, the kinetic energy per unit mass in the dark-energy case is ${\rm KE}/M\propto \ell^2$ (equation~(\ref{v11})), whereas in the absence of dark energy,  ${\rm KE}/M\propto M^{2/3}$ (equation~(\ref{v3})). This is the reason that the solutions of \S~\ref{sec3} are not reducible to those shown in \S~\ref{sec2} as $\ell\to 0$.

\section*{Acknowledgments}

We appreciate the referee's effort to improve this work. NASA support over the years is also gratefully acknowledged.

\label{lastpage}

\end{document}